\documentclass[conference]{IEEEtran}
\IEEEoverridecommandlockouts
\usepackage{cite}
\usepackage{amsmath,amssymb,amsfonts}
\usepackage{algorithmic}
\usepackage{graphicx}
\usepackage{textcomp}
\usepackage{xcolor}
\usepackage{makecell}
\usepackage{comment}
\usepackage{flushend}
\def\BibTeX{{\rm B\kern-.05em{\sc i\kern-.025em b}\kern-.08em
    T\kern-.1667em\lower.7ex\hbox{E}\kern-.125emX}}

\begin{document}

\title{A multi-module silicon-on-insulator chip assembly containing quantum dots and cryogenic radio-frequency readout electronics}

\author{\IEEEauthorblockN{David J. Ibberson\IEEEauthorrefmark{1},
James Kirkman\IEEEauthorrefmark{1}, John J. L. Morton\IEEEauthorrefmark{1}\IEEEauthorrefmark{2},
M. Fernando Gonzalez-Zalba\IEEEauthorrefmark{1}, Alberto Gomez-Saiz\IEEEauthorrefmark{1}}
\IEEEauthorblockA{\IEEEauthorrefmark{1}Quatum Motion, 9 Sterling Way, London, N7 9HJ, United Kingdom}
\IEEEauthorblockA{\IEEEauthorrefmark{2}London Centre for Nanotechnology, UCL, London, WC1H 0AH, United Kingdom}
Email: david@quantummotion.tech, alberto@quantummotion.tech
}

\maketitle

\begin{abstract}
Quantum processing units will be modules of larger information processing systems containing also digital and analog electronics modules. Silicon-based quantum computing offers the enticing opportunity to manufacture all the modules using the same technology platform. Here, we present a cryogenic multi-module assembly for multiplexed readout of silicon quantum devices where all modules have been fabricated using the same fully-depleted silicon-on-insulator (FDSOI) CMOS process. The assembly is constituted by three chiplets: (i) a low-noise amplifier (LNA), (ii) a single-pole eight-throw switch (SP8T), and (iii) a silicon quantum dot (QD) array. We integrate the chiplets into modules and show respectively, (i) a peak gain over 35~dB with a 3-dB bandwidth from 709~MHz to 827~MHz and an average noise temperature of 6.2~K (minimum 4.2~K), (ii) an insertion loss smaller than 1.1~dB and a noise temperature less than 1.1~K over the 0-2~GHz range, and (iii) single-electron box (SEB) charge sensors. Finally, we combine all modules into a single demonstration showing time-domain radio-frequency multiplexing of two SEBs paving the way to an all-silicon quantum computing system. 
\end{abstract}


\section*{Introduction}
Quantum computing hardware based on spins in silicon QDs is a promising approach towards a scalable quantum computing system. Particularly, silicon spin qubits have been operated and read out with a precision above the threshold to perform quantum error correction~\cite{bib1}. In terms of scaling, quantum processing units of up to 6 qubits have been fabricated using experimental processes~\cite{bib2, bib3, bib4} but substantial further scaling may be subject to exploiting silicon’s CMOS manufacturing infrastructure~\cite{bib5} enabling integration with classical cryo-electronics to form a compact quantum computing system~\cite{bib6,bib7}. To this purpose, demonstrating QD devices as well as cryo-electronics modules in an industry standard process is a key step towards a fully-fledged quantum processor. Here, we demonstrate a multi-module assembly containing two cryo-electronic modules -- a RF LNA and a RF Switch -- as well as a quantum module -- an array of QD-based sensing devices -- all manufactured using GlobalFoundries 22-nm FDSOI technology (22FDX). Our previous work had developed cryogenic models in this technology~\cite{bib15} which are used to aid in the circuit design. We combine all modules into a time-domain multiple access Radio-Frequency (RF) multiplexing demonstration of SEB charge detection showing key developmental steps towards an all silicon quantum processing system. 

\begin{figure}
    \centering
    \includegraphics[width=0.85\linewidth]{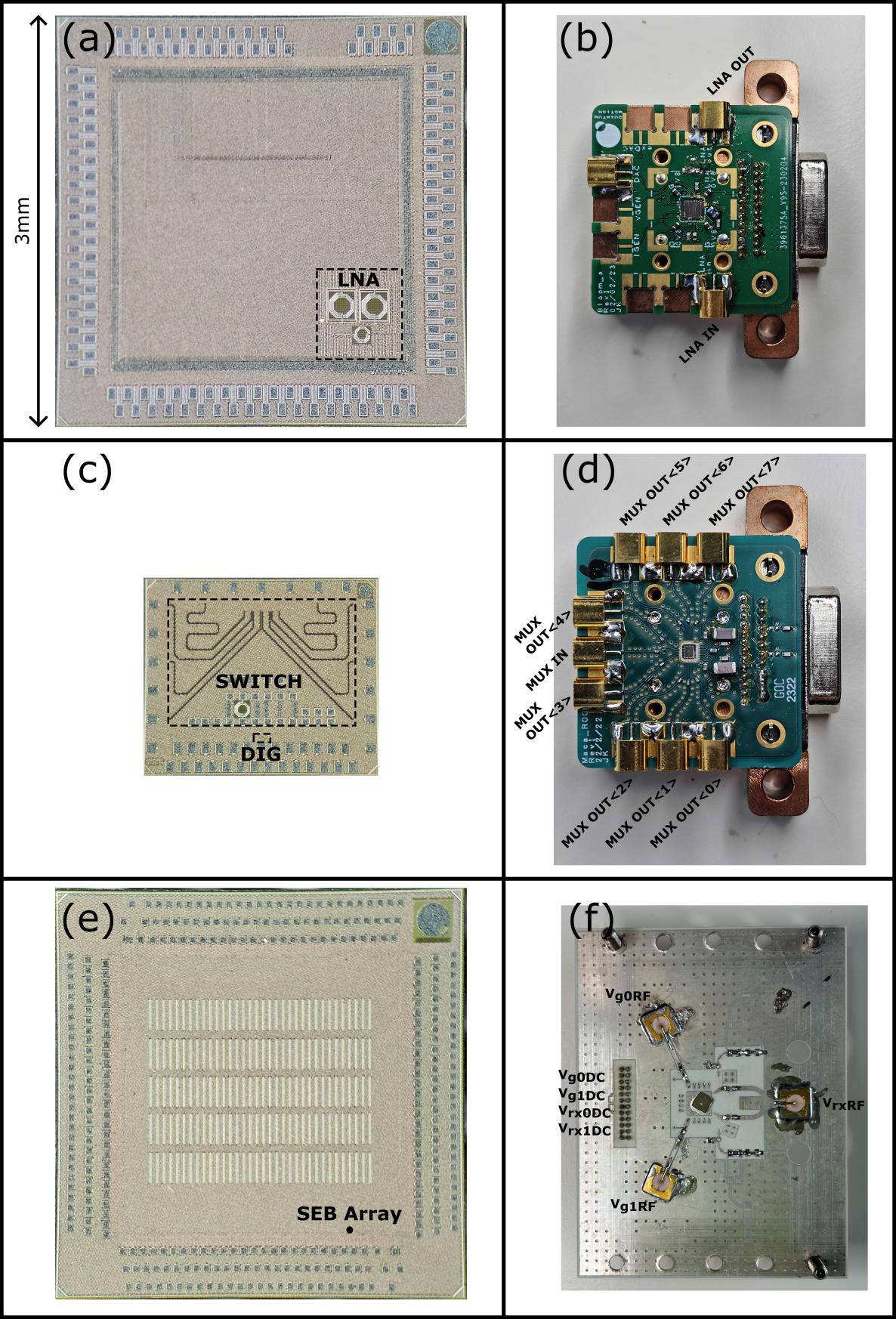}
    \caption{Chiplets and Modules. (a) LNA chiplet micrograph. (b) LNA module photo. (c) Switch chiplet micrograph. (d) Switch module photo. (e) SEB array chiplet micrograph. (f) SEB array module photo.}
    \label{fig:photos}
\end{figure}

\section*{Cryogenic low-noise amplifier}

\begin{figure}
    \centering
    \includegraphics[width=1\linewidth]{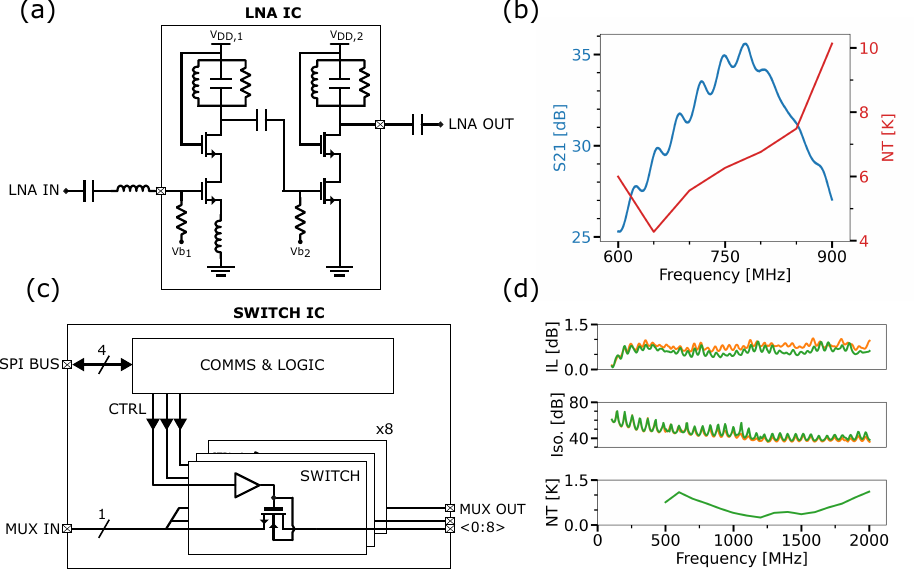}
    \caption{LNA and Switch ICs (a) LNA circuit schematic. (b) LNA gain (blue) and Noise Figure (red). (c) Switch schematic. (d) Insertion Loss (top), Cross-channel isolation (middle) and Noise Temperature (bottom) vs frequency. Green (orange) trace typical (worst) case. Typical Noise Temperature (bottom). Measurements performed at 4~K ambient temperature. }
    \label{fig:IC}
\end{figure}

To provide cryogenic amplification, we designed a two stage RF LNA IC. Each stage uses a cascoded topology with a passive resonator as load. The internal bias unit and external positive supply of each stage are independent to ensure stability. The first stage uses inductive source degeneration to provide narrow-band input matching. Both stages use thin oxide flipped-well NMOS devices which have shown excellent RF performance in the deep-cryogenic regime~\cite{bib9}. Figure~\ref{fig:IC}(a) shows the simplified schematic of the LNA IC with additional external passives for input and output matching. We integrated the IC and passives in a PCB module (Fig.~\ref{fig:photos}(a) and (b)), and characterized it at 4~K ambient temperature. Figure ~\ref{fig:IC}(b) shows the measurement results. We show a peak S$_{21}$ greater than 35~dB at ~780~MHz and a 3dB bandwidth from 709~MHz to 827~MHz, an input-referred average NT of 6.2~K over the LNA bandwidth, and a minimum NT of 4.2~K at 650~MHz, see Fig.~\ref{fig:IC}(b). Note that the noise match happens below the LNA bandwidth due to not accurately modelled changes in the electrical characteristics of the devices in the deep-cryogenic regime. The measured standby power consumption of the module is 36~mW.

\section*{Cryogenic SP8T switch}

To provide cryogenic time-domain RF multiplexing, we designed a SP8T RF switch IC. The IC is comprised of: a digital block, and a RF switch block. The digital block provides external communication using the SPI protocol and generates the control signals for the RF block. The RF block contains eight series-only switching units. The switching units uses thick oxide flipped-well NMOS devices operated in dynamic threshold-voltage mode. Figure~\ref{fig:IC}(c) shows the block diagram of the IC including a simplified schematic of the RF block. We integrated the chiplet in a PCB module (Fig.~\ref{fig:photos}(c) and (d)), and characterized it in the 0-2~GHz input frequency range and at 4K ambient temperature (Fig. ~\ref{fig:IC}(d)). We show for all channels an insertion loss of less than 1.1~dB, a cross-channel isolation better than 35~dB and a NT smaller than 1.1~K. The measured standby power consumption of the module is below the measurement resolution of 100~nW, and within the cooling budget of the mixing chamber~\cite{bib8}. We functionally validated the module at mK ambient temperature and expect its performance to be comparable to the 4~K measurements~\cite{bib16}.

\section*{Single-electron box charge sensor}

\begin{figure}
    \centering
    \includegraphics[width=0.85\linewidth]{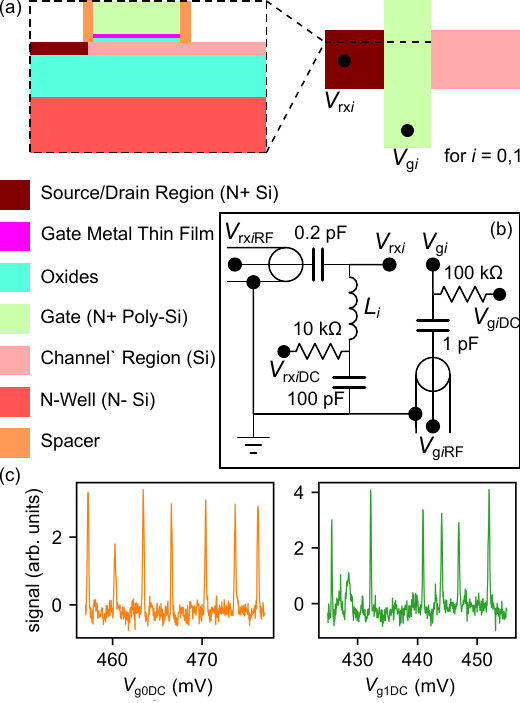}
    \caption{Single-electron box chiplet. (a) Schematic views of the SEB device: (top left) cross-section along the silicon channel direction and (top right) top view. A QD forms in the silicon directly under the gate electrode and is tunnel coupled to the ohmic contact. $V_{\text{rx}i}$ refers to the ohmic contact potential and $V_{\text{g}i}$ to the gate potential. The schematic elements are indicated with colored labels (bottom left). (b) Schematic of the matching network and components used in the signal delivery lines. $V_\text{rx(g)DC}$ refer to quasi-static voltages and $V_\text{rx(g)RF}$ to radio-frequency voltages. (c) Demodulated transmitted signal as a function of gate voltage showing charge oscillations for SEB:0,1; orange and green traces, respectively.}
    \label{fig:SEB}
\end{figure}

We present results on SEBs charge detectors, (SEB$_i$ for $i=0,1$). The devices are implemented in a narrow channel multi-gate transistor, see Fig.~\ref{fig:SEB}(a). When a positive potential, near the threshold voltage, is applied to a gate (G$_i$) placed next to an ohmic contact (RX$_i$), a QD tunnel-coupled to an electron reservoir forms, i.e a SEB. We probe the impedance of the device using RF transmission techniques by applying an RF excitation to G$_i$ and collecting the transmitted signal on the RX$_i$ port where a sub-1 GHz L-shape high-pass matching resonant network of center frequency $f_i$ is placed~\cite{bib10}. We refer to these matching networks as MN:i, see Fig.~\ref{fig:SEB}(b). When electrons cyclically tunnel between the QD and the reservoir due to the RF excitation, a change in the transmission through the device occurs. We perform the measurements on two SEBs at mK ambient temperature as a function of the respective gate voltage, see orange and green traces in Fig.~\ref{fig:SEB}(c). Such sharp change in transmission can be used for sensitive charge detection of nearby QDs or qubits~\cite{bib11}.  The linewidth of these transitions can be used to provide an upper bound of the electron temperature of the SEBs~\cite{bib12}, for which we find 360~mK. We integrated the chiplet in a PCB module (Fig.~\ref{fig:photos}(e) and (f)). 

\begin{figure}
    \centering
    \includegraphics[width=0.9\linewidth]{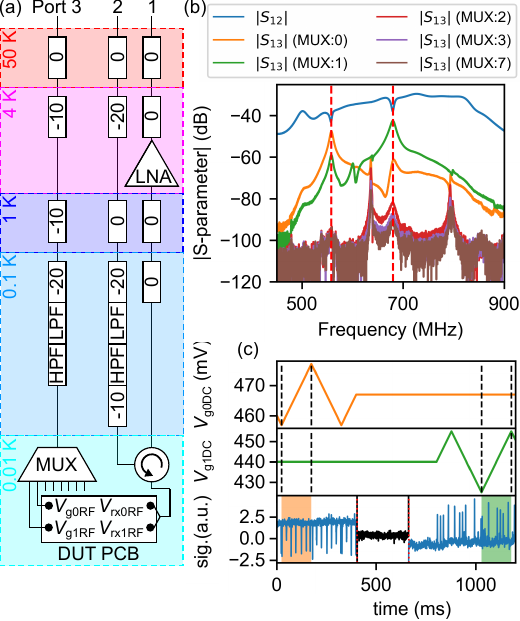}
    \caption{Multi-module assembly and cryogenic setup. (a) Setup inside a dilution refrigerator. Rectangles indicate attenuation or low- or high-pass filtering stages. (b) S-parameter characterisation. The different traces correspond to the $|S_{ij}|$ (MUX:$k$) where $i(j)$ refers here to the output(input) port and $k$ to the active MUX output channel. (c) Time-division multiplexing. Time-dependent voltage applied to gate 0 (top) and 1 (middle). Time trace of the demodulated signal (bottom).}
    \label{fig:multi}
\end{figure}

\section*{Multi-module cryogenic assembly}

Next, we move on to the cryogenic characterisation of the full assembly, see Fig.~\ref{fig:multi}(a). We send an attenuated RF signal to the switch module which acts as a multiplexer (MUX) placed at mixing chamber of a dilution refrigerator with base temperature of 20~mK. The MUX distributes the signal on-demand through channels MUX:0 and MUX:1 to the SEB chiplet also placed at the mixing chamber plate. Then the transmitted signal through each SEBs is impedance-matched at the output using the $LC$ networks. The signal is then amplified by the LNA and base-band IQ demodulated at room temperature. 

We now characterize the S-parameters of the assembly, see Fig.~\ref{fig:multi}(b). First, in blue, we plot a $S_{12}$ measurement, equivalent to a RF reflectometry test, showing two sharp resonances at the frequency of the $LC$ resonators (red dashed lines) from which we determine $f_i$. We now characterize the transmission through the assembly, $S_{13}$, including the MUX. In orange(green), we activate MUX:0(1) and show the transmission primarily happens through MN:0(1) with an isolation to MN:1(0) of $>13$~dB. Additionally, we show measurements when MUX:2,3,7 are active (red, purple and brown lines) indicating a $> 39$~dB internal MUX isolation. Cross-coupling at the SEB chiplet contributes to the difference in isolation. Finally, in Fig.~\ref{fig:multi}(c), we combine the three modules into a demonstration of multi-frequency time-division multiplexing. We continuously send two RF tones at $f_0=559$~MHz and $f_1=681$~MHz. First, for $t<400$~ms, we select MUX:0 and detect the charge oscillations as a function of gate voltage in SEB$_0$ (see the blue trace in the bottom panel). During this time period, we ramp the gate voltage of SEB$_0$ (top panel) to change the charge state of the SEB, while the gate voltage on SEB$_1$ remains static (middle panel). For $400\text{~ms}<t<660\text{~ms~}$, we deselect the MUX. For $t>660$~ms, we select MUX:1 and detect charge oscillations in SEB$_1$ swapping the voltage profiles on the SEB gates. We record the data while the corresponding gate voltage is being ramped up (time windows between the vertical dashed lines). Within the time resolution of our measurements which is below the inverse bandwidth of the resonators ($\sim3$~MHz), we do not observe any transient after the switching events. 

Next, we benchmark the sensitivity of the assembly in terms of the power signal-to-noise ratio (SNR). We define the signal as the square of the voltage amplitude of a SEB oscillation as measured in the IQ plane and the noise as the square of the average standard deviation of the voltage levels at the top and bottom of the oscillation. We find a SNR of 140 at 10~$\mu$s integration time, corresponding to an integration time for SNR of 1 of 70~ns assuming a white noise profile dominated by the LNA. The result indicates that high-fidelity readout of silicon spin qubits could be achieved in timescales well below 10~$\mu$s, a results that compares favorably to previous demonstrations~\cite{bib11,bib22}. More particularly, a readout fidelity well above 99\%, which fulfills the readout requirements to implement a fault-tolerant quantum computer,  could be achieved within a 1~$\mu$s integration time. This time is comparable to the typical surface code cycle in silicon spin qubits~\cite{bib13}. 

\begin{table*}
    \centering
    \caption{Performance summary and comparison cryo-LNA}
    \begin{tabular}{|c|c|c|c|c|}
    \hline
    \textbf{Ref.} & \textbf{This Work}& \textbf{IMS'23~\cite{bib17}}& \textbf{MWCL'22~\cite{bib18}}& \textbf{JSSC'18~\cite{bib19}} \\
    \hline
    Technology & \makecell{22-nm \\ FDSOI CMOS} & \makecell{40-nm \\ Bulk CMOS} & \makecell{65-nm \\ Bulk CMOS} & \makecell{160-nm \\ Bulk CMOS} \\
    \hline
    Topology & Cascode CS & \makecell{Inverted-based input \\ + Noise-Canceling} & \makecell{Folded Cascode \\ CS} & Noise-Canceling \\
    \hline
     Meas. Temp.~(K) & 4 & 4 & 20 & 4.2 \\
    \hline
    Freq.~(GHz) & $\text{0.71}\sim\text{0.82}$ & $\text{0.01}\sim\text{3}$ & $\text{0.9}\sim\text{1.8}$ & $\text{0.1}\sim\text{0.5}$ \\
    \hline
    $S_{21}$~(dB) & 35 & 29 & 37.2 & 57 \\
    \hline
    $S_{11}$~(dB) & $<-10$ & $-11.2\sim-8.3$ & $<-10$ & $-7\sim-3$ \\
    \hline
    NT~(K) & $\text{4.2}\sim\text{7.1}$ & - & $\text{2.0}\sim\text{8.8}$ & $\text{6.8}\sim\text{62.7}$ \\
    \hline
    Power~(mW) & 36 & 19.4 & 125 & 91 \\
    \hline
    Area~($\text{mm}^2$) & 0.349 & 0.018$^{\mathrm{a}}$ & 1$^{\mathrm{a}}$ & 0.249$^{\mathrm{a}}$ \\
    \hline
    \multicolumn{5}{l}{$^{\mathrm{a}}$Estimated from figures.}
    \end{tabular}
    \label{tabLNA}
\end{table*}

\begin{table*}
\begin{minipage}[t]{\columnwidth}
    \centering
    \caption{Performance summary and comparison cryo-Switch}
    \begin{tabular}{|c|c|c|c|}
        \hline
        \textbf{Ref.} & \textbf{This Work}& \textbf{\makecell{Nat. \\ Elec.'23~\cite{bib20}}}& \textbf{IMS'22~\cite{bib21}} \\
        \hline
        Technology & \makecell{22-nm \\ FDSOI CMOS} & \makecell{28-nm \\ Bulk CMOS} & \makecell{22-nm \\ FDSOI CMOS} \\
        \hline
        Configuration & SP8T & SP4T & SPST \\
        \hline
        Meas. Temp.~(K) & 4 & 4 & 2 \\
        \hline
        Freq.~(GHz) & $\text{0.01}\sim\text{2}$ & $\text{4}\sim\text{8}$ & $\text{DC}\sim\text{70}$ \\
        \hline
        IL~(dB) & $<\text{1.1}$ & $<\text{5}$ & $<\text{2.3}$ \\
        \hline
        Iso.~(dB) & $>\text{35}$ & $>\text{30}$ & $<\text{25.3}$ \\
        \hline
        RL~(dB) & $>10$ & $8.3\sim11.2$ & $>11.5$ \\
        \hline
        NT~(K) & $<\text{1.1}$ & 0.15$^{\mathrm{a}}$ & - \\
        \hline
        Area~($\text{mm}^2$) & 2.01 & 1$^{\mathrm{b}}$ & 0.00016$^{\mathrm{c}}$ \\
        \hline
        \multicolumn{4}{l}{$^{\mathrm{a}}$At 6~GHz.} \\
        \multicolumn{4}{l}{$^{\mathrm{b}}$Estimated from figures.} \\
        \multicolumn{4}{l}{$^{\mathrm{c}}$Core area only.}
    \end{tabular}
    \label{tabSW}
\end{minipage}\hfill 
\begin{minipage}[t]{\columnwidth}
    \centering
    \caption{Performance summary and comparison SEB readout}
    \begin{tabular}{|c|c|c|c|}
        \hline
        \textbf{Ref.} & \textbf{This Work}& \textbf{Nat.'22~\cite{bib22}}& \textbf{PRX'23~\cite{bib11}} \\
        \hline
        Technology & \makecell{22-nm \\ FDSOI CMOS} & Si/SiGe & \makecell{Si-SOI \\ nanowire} \\
        \hline
        \makecell{Resonator \\ Type} & Multi-module & On-chip & Multi-module \\
        \hline
        \makecell{Channel \\ Management} & TDMA & N.A. & N.A. \\
        \hline
        t$_{\text{min}}$~(ns) & 70 & 1.8 & 170 \\
        \hline
    \end{tabular}
    \label{tabSEB}
  \end{minipage}
\end{table*}

\section*{Conclusions}

We have demonstrated a multi-module assembly for the multiplexed RF readout of silicon QD devices where all chiplets have been manufactured using the GlobalFoundries 22FDX process. In the demonstration, we include a RF switch for the delivery of time-division multiplexed radio-frequency signals to an array of single-electron , whose transmitted signal is 50-ohm matched and amplified using a cryogenic low-noise amplifier. We benchmark the sensitivity of the assembly and find a power SNR of $140$ for a readout time of 10~$\mu$s, approaching state-of-the-art values~\cite{bib14}. We use resonators at different frequencies to demonstrate the capability of the LNA to support multiple readout channels via frequency-division multiplexing. Our results highlight the potential of CMOS technology towards the realization of a quantum computing system containing not only quantum devices but also cryogenic analog and digital electronics.

\section*{Acknowledgments}

The authors acknowledge support from Jonathan Warren of Quantum Motion. M.~F.~G.-Z. acknowledges a UKRI Future Leaders Fellowship [MR/V023284/1].

\end{document}